\documentclass[11pt, a4paper]{article}

\title{Fraud Analytics: A Decade of Research\\
	   Organizing Challenges and Solutions in the Field}

\date{November, 2022}

\author{
        Christopher Bockel-Rickermann, KU Leuven
        \thanks{Research Centre for Information Systems Engineering (LIRIS)}
        \and
        Tim Verdonck, University of Antwerp
        \thanks{Department of Mathematics}
        \and
        Wouter Verbeke, KU Leuven\footnotemark[1]
        }
        
\usepackage[T1]{fontenc} 

\usepackage[english]{babel}
	
\usepackage{csquotes} 
\usepackage[style=apa, citestyle=apa, backend=biber]{biblatex} 
	
\usepackage{amsmath}
\usepackage{amssymb}
\usepackage{amsthm}
\usepackage{amscd}
\usepackage{amsfonts}
\usepackage{bm}         
	
\usepackage{graphicx}   
\usepackage{courier}    
\usepackage{enumitem}   
\usepackage{booktabs}   
\usepackage{fancyhdr}   
\usepackage{longtable}  
\usepackage{subfigure}  
\usepackage[usenames,dvipsnames]{xcolor}    
\usepackage{array}		
\usepackage{wasysym}	
\usepackage{fancybox}	
\usepackage{rotating}	
\usepackage{diagbox} 	
\usepackage{multirow}	
\usepackage{adjustbox}	
\usepackage{ragged2e}	
\usepackage[font=small,labelfont=bf,justification=justified,singlelinecheck=no,parskip=5pt,margin=0.2cm]{caption} 
\usepackage{arydshln}	
\usepackage{setspace}	
\usepackage[left=72pt, right=72pt, right=72pt, top=72pt]{geometry}  
\usepackage{titlesec}	
\usepackage{lipsum}     
\usepackage{float}		
	
\newtheorem{definition}{Definition}

\titleformat{\subsubsection}{}{\thesubsubsection}{1em}{\itshape}

\nocite{*}

\definecolor{KULblue}{RGB}{29,141,176} 

\newcommand\sbullet[1][.5]{\mathbin{\vcenter{\hbox{\scalebox{#1}{$\bullet$}}}}}

\makeatletter
\def\nobreakhline{%
	\noalign{\ifnum0=`}\fi
	\penalty\@M
	\futurelet\@let@token\LT@@nobreakhline}
\def\LT@@nobreakhline{
	\ifx\@let@token\hline
	\global\let\@gtempa\@gobble
	\gdef\LT@sep{\penalty\@M\vskip\doublerulesep}
	\else
	\global\let\@gtempa\@empty
	\gdef\LT@sep{\penalty\@M\vskip-\arrayrulewidth}
	\fi
	\ifnum0=`{\fi}%
	\multispan\LT@cols
	\unskip\leaders\hrule\@height\arrayrulewidth\hfill\cr
	\noalign{\LT@sep}%
	\multispan\LT@cols
	\unskip\leaders\hrule\@height\arrayrulewidth\hfill\cr
	\noalign{\penalty\@M}%
	\@gtempa}
\makeatother

\begin{document}


\setcounter{page}{1} \renewcommand{\thepage}{\arabic{page}} 

\maketitle

\begin{abstract}
    The literature on fraud analytics and fraud detection has seen a substantial increase in output in the past decade. This has led to a wide range of research topics and overall little organization of the many aspects of fraud analytical research. The focus of academics ranges from identifying fraudulent credit card payments to spotting illegitimate insurance claims. In addition, there is a wide range of methods and research objectives. This paper aims to provide an overview of fraud analytics in research and aims to more narrowly organize the discipline and its many subfields.  We analyze a sample of almost 300 records on fraud analytics published between 2011 and 2020. In a systematic way, we identify the most prominent domains of application, challenges faced, performance metrics, and methods used. In addition, we build a framework for fraud analytical methods and propose a keywording strategy for future research. One of the key challenges in fraud analytics is access to public datasets. To further aid the community, we provide eight requirements for suitable data sets in research motivated by our research. We structure our sample of the literature in an online database. The database is available online for fellow researchers to investigate and potentially build upon.
\end{abstract}
	
\section{Introduction\label{sec:Intro}}

By the end of 2021, a search for ``\textit{fraud detection}'' and ``\textit{fraud analytics}'' on the Scopus database of academic records yielded almost 4,000 entries. A number that shows the popularity of the associated research field. Further, according to our own analyses, the above literature has seen approximately 20\% growth in output year-over-year from 2011 to 2020.

To date, fraud detection ranges from identifying nontechnical losses in electricity grids to finding unlawful tax statements. Differences in the nature of scientific publications do not stop there. Researchers and practitioners utilize and produce papers of different types, domains and research questions, all summarized under the same terminology. To date, there has been little research across domains of applications, leading to the assumption that the broader scientific field of fraud detection and fraud analytics is lacking organization. With the strong growth in output, this issue is expected to persist and potentially worsen in the near-term future. For an applied science that has such tangible real-world applications \parencite[][]{Baesens.2015}, the implications are severe and range from redundancy of work to risking a divergence of academia and practice \parencite[][]{RymanTubb.2018}. Hence, organizing literature and research efforts should be an imperative. Yet, to our knowledge, little has been done to establish it.

This paper will focus on data-driven fraud detection research, further referred to by ``\textit{fraud analytics}''. We intend to provide a holistic overview of the field and to organize its many branches. This way, we provide guidance for readers and future research alike. Different from existing surveys and reviews, we will not bound ourselves to a certain domain, technical aspect or methodology, but will focus on the general topic of ``fraud analytics''. We elicit trends in domain popularity, research questions and key issues in the analytical detection of fraud. Moreover, we seek to identify data-driven methods and their most popular applications in the field and enable a sound gap analysis of possible future research questions. We intend to provide the community with a taxonomy to further classify and guide future research by proposing a framework that fits the fraud analytical methods we find in the literature.

To accomplish our goal, we assessed the literature on fraud analytics via a systematic search and analysis of the \textit{Scopus} and \textit{Web of Science} databases. We gathered a sample of the fraud analytical literature by selecting almost 300 papers dealing with or relating to the field published between 2011 and 2020. We clustered papers by their type, domain, challenges faced, methods used and performance metrics applied. Thereby, we distinguish our paper from previous surveys and reviews by both using a domain-unspecific body of literature and by providing organization on top of analysis. Previous surveys and reviews have so far been focused on specific domains and methods, typically restricting themselves to specific use cases. In addition to the obtained insights, we present an approach to cluster fraud analytical research, as well as a taxonomy to guide future researchers in positioning their work.

To our knowledge, the above approach is novel to the field of fraud analytics and adds value in several dimensions:
\begin{itemize}
	\item We provide a detailed picture of the fraud analytical research in the last decade and thereby review an important part of the academic output created on the topic.
	\item We review data-driven methods used in fraud detection and describe the current state-of-the-art from a majority vote perspective.
	\item We derive the influential challenges and issues faced in the fraud analytical literature and map methods used to tackle those.
	\item We find gaps in the literature and enable researchers and practitioners to actively bridge between research and practice.
	\item We propose a taxonomy to classify existing and future research on fraud analytics.
\end{itemize}

The remainder of our paper is organized as follows: In Section~\ref{sec:LitReview}, we provide an overview of the existing reviews and surveys of the fraud analytical literature and position our work. Section~\ref{sec:Method} will detail our methodology, i.e., our search strategy and clustering framework to derive the results presented in Section~\ref{sec:Results}. In Section~\ref{sec:ResearchDirs}, we further discuss insights from assessing our results, highlighting gaps and potential future research directions. In Section~\ref{sec:Database}, we introduce a keyword taxonomy to facilitate researchers in precisely situating their work within the body of literature on fraud analytics. Additionally, we present an online database in which we publish our full results. Finally, we conclude the paper in Section~\ref{sec:Conclusion}.

\section{Related Works\label{sec:LitReview}}

We find several reviews and surveys on fraud analytics in the existing literature. As indicated in Section~\ref{sec:Intro}, many of these studies focus on specific domains \parencite[][]{Ahmed.2016, Wang.2016, Mehrban.2020} or methodologies \parencite[][]{Johnson.2019c, Zliobaite.2016, Pourhabibi.2020}.

With respect to our approach and aim, we find a smaller number of closely related studies: \textcite{Ngai.2011} perform a survey of analytical methods for fraud detection in financial domains: that is, bank fraud, insurance fraud and securities fraud. Moreover, they propose a framework for the classification of fraud detection methods, taking a step towards organizing and structuring research in the field. The clusters of methods are classification, clustering, outlier detection, prediction, regression and visualization. Our research seamlessly generalizes this work by not only extending the list of possible applications but also proposing a more detailed clustering framework for methods (cf. Section~\ref{sec:Method}). \textcite{Phua.2010} perform a review of fraud analytical research from 2000 to 2010. They analyse 51 studies on fraud analytics and classify them by their domain of application, performance metrics and methods used. The authors differentiate between supervised, unsupervised and semisupervised methods, establishing a framework that we will incorporate in our own methodology (cf. Section~\ref{sec:Method}). Lacking, in comparison to our research, is the notion of challenges encountered in the literature and the linkage between dimensions. An earlier collection of fraud analytical domains is found in \textcite{Bolton.2002}. Finally, \textcite{RymanTubb.2018} surveyed challenges and methods in payment fraud detection. The authors do not survey across a wider array of domains, yet compare the performance of methods in a meta analysis of records. \textcite{RymanTubb.2018} aim explicitly to bridge between research and industry. We agree with the authors on the importance of doing so and hope that we can further stipulate bridging between industry and academia.

The reader will note that our work methodologically positions between \textcite{Ngai.2011} and \textcite{Phua.2010}. The vast growth of fraud analytical literature since its publication adds to the importance and value of our work.

\section{Methodology\label{sec:Method}}

\subsection{A Systematic Review of Fraud Analytical Literature\label{sec:Method_SysReview}}

In the following we will detail the methodology used to sample and organize the fraud analytical literature. Our approach is motivated by and based on the scientific literature reviews proposed by \textcite{Tranfield.2003}. We follow five steps, namely, ``Scope identification", ``Literature search", ``Record selection", ``Reading and analysis" and ``Synthesis". Each of the steps is defined carefully to be suited to approach the fraud analytical literature as holistically as possible:

\subsubsection*{Step 1 - Scope Identification:}

We focus on fraud analytics. Further, we do not limit ourselves to a subdomain as typically seen in other reviews and surveys on fraud analytics (cf. Section~\ref{sec:LitReview}). As discussed in Section~\ref{sec:Intro}, we focus on quantitative fraud detection, that is ``\textit{fraud analytics}". We define fraud analytics in the following way:

\begin{definition}
\label{def:fraud_analytics}
\textit{``Fraud analytics" refers to the use of data-driven methods to discover, recognize and detect fraudulent activities in sets or streams of data. ``Fraud analytics" is developing and presenting a toolbox for practitioners to engage in the detection of fraud in their respective domain.}
\end{definition}

We target addressing the field holistically and try to impose as few constraints on its scope as possible. A similar scope has so far only been provided by compendium-type literature, such as \textcite{Baesens.2015}. Studies with different perspectives such as law \parencite[e.g., ][]{Ramamoorti.2008} or psychology \parencite[e.g., ][]{Kingston.2004}, are beyond the scope.

\subsubsection*{Step 2 - Literature Search:}

We searched the literature on both the Scopus and the Web of Science (WoS) databases. Combined, these databases have a high degree of completeness and include a wide selection of outlets and records. Both databases are widely respected and used in research \parencite[][]{Burnham.2006, Mongeon.2016}.
Therefore, we believe that this selection, though a limitation, is sufficient for the purpose of our research. Furthermore, we restricted our search to articles in scientific journals and conference proceedings so that records were comparable with respect to audience and style.

\subsubsection*{Step 3 - Record Selection:}

From the resulting literature, we investigate all articles written in English and published between 2011 and 2020. The latter gives our review a degree of contemporaneity and adds seamlessly to the aforementioned studies from \textcite{Ngai.2011} and  \textcite{Phua.2010}. Note that, due to the strong growth in academic output, the time interval accounts for the majority of literature in the field\footnote{Literature from 2021 onward is left out of scope due to further filtering steps described in the text below.}.

\begin{figure}[ht]
	\centering
	\footnotesize
	\includegraphics[width=0.6\textwidth]{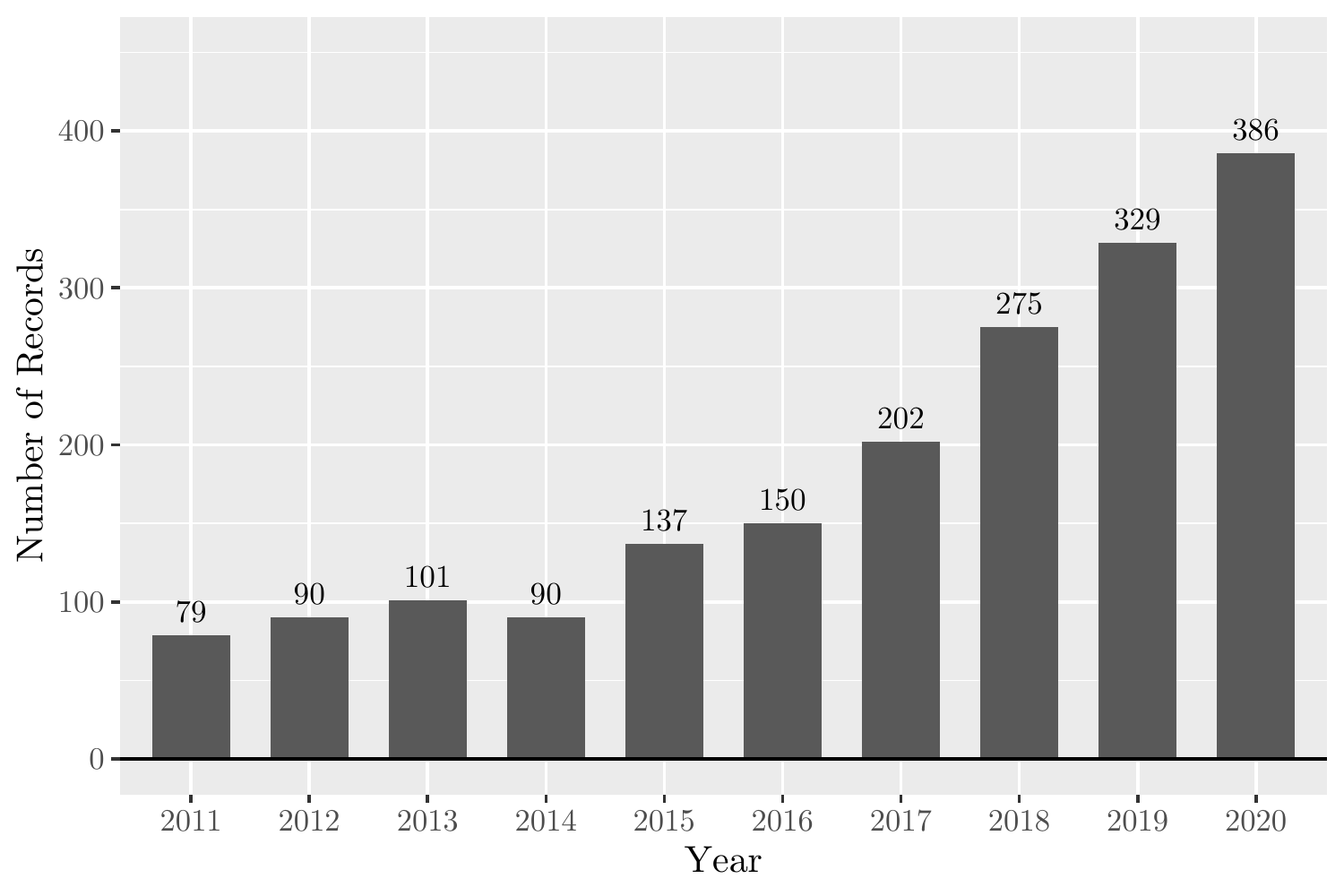}
	\caption[Records per year]{Records per year}
	\label{fig:Records_per_year}
\end{figure}

The approach yielded 1,847 records\footnote{The literature search and data acquisition were performed in September 2021.}. A distribution across years is shown in Figure~\ref{fig:Records_per_year}. We can easily spot the strong growth in output per year as mentioned before in Section~\ref{sec:Intro}. To facilitate analysis we further reduce the sample size by retaining only the top 15\% most-cited papers per year. This step aims to yield a sample of well-received academic output, where ``well-received" is based on the number of citations of the respective record. Likewise, this step increases the average quality of the records. 

We do not base the selection of papers on impact factors, since impact factors indicate the general relevance the journal rather than the relevance of a particular paper. Citations, in contrast, reflect both the size of a record's audience and its reception. Additionally, we note several criticisms of the use of impact factors for sampling literature \parencite[][]{Seglen.1997, Dong.2005, Simons.2008}. We are aware of the scepticism towards citations as a mean to select literature. However, as we intend to organize across sub-fields of fraud analytics, citations enable us to reduce the implicit bias that a discretionary selection of records would impose, that is, a focus on a selected few popular domains or fraud analytical methods.

The entire search strategy is depicted in Table~\ref{tbl:Search_streatgy}. Search queries can be found in Table~\ref{tbl:SearchQueries} in the Appendix.

\begin{table}[t] 
	\caption[Search strategy and number of records]{Search strategy and number of records}
	\label{tbl:Search_streatgy}
	\centering
	\begin{adjustbox}{max width=451pt}
		\renewcommand{\arraystretch}{0.8}
		\begin{tabular}{@{\extracolsep{2pt}}lllcc} 
			\\[-1.0ex]\hline 
			\hline \\[-1.0ex] 
			\multirow{2}{*}{Filtering Step} & \multicolumn{2}{l}{\multirow{2}{*}{Decision}} & \multicolumn{2}{c}{Number of records} 	\\ [-0.1ex] \cdashline{4-5} \\[-1.0ex]
											&&												& Scopus & Web of Science		\\
			\hline \\[-1.0ex]			
			\multirow{2}{*}{\textbf{Database selection}} & \multicolumn{2}{l}{$\sbullet[.75]$ Scopus} & \multirow{2}{*}{\textit{Not meaningful}} & \multirow{2}{*}{\textit{Not meaningful}}\\
												& \multicolumn{2}{l}{$\sbullet[.75]$ Web of Science} \\[0.1ex] \hdashline \\ [-1.0ex]
			\textbf{1. Topic selection} & \multicolumn{2}{l}{$\sbullet[.75]$ ``Fraud detection" $\parallel$ ``Fraud analytics"} & 3,918 & 2,593 \\[0.1ex] \hdashline \\ [-1.0ex]
			\textbf{2. Topic selection} & \multicolumn{2}{l}{$\sbullet[.75]$ ``Data"} & 2,760 & 1,653 \\[0.1ex] \hdashline \\ [-1.0ex]
			\textbf{Years} & \multicolumn{2}{l}{$\sbullet[.75]$ 2011 - 2020} & 1,970 & 1,242 \\[0.1ex] \hdashline \\ [-1.0ex]
			\multirow{2}{*}{\textbf{Outlets}} & \multicolumn{2}{l}{$\sbullet[.75]$ Journals} & \multirow{2}{*}{1,788} & \multirow{2}{*}{1,204}\\
			& \multicolumn{2}{l}{$\sbullet[.75]$ Conference Proceedings} \\[0.1ex] \hdashline \\ [-1.0ex]
			\multirow{3}{*}{\textbf{Fields}} & $\sbullet[.75]$ Business & $\sbullet[.75]$ Computer Sci. & \multirow{3}{*}{1,686} & \multirow{3}{*}{1,108}\\
			& $\sbullet[.75]$ Economics \qquad \quad \quad & $\sbullet[.75]$ Engineering \\
			& $\sbullet[.75]$ Mathematics \\[0.1ex] \hdashline \\ [-1.0ex]
			\textbf{Language} & \multicolumn{2}{l}{$\sbullet[.75]$ English} & 1,653 & 1,094 \\[0.1ex] \hdashline \\ [-1.0ex]
			\textbf{Duplicates} & \multicolumn{2}{l}{$\sbullet[.75]$ Excl. double counting} & \multicolumn{2}{c}{1,847} \\[0.1ex] \hdashline \\ [-1.0ex]
			\textbf{Citations} & \multicolumn{2}{l}{$\sbullet[.75]$ Top 15\% most cited per year} & \multicolumn{2}{c}{294} \\[0.1ex] \hdashline \\ [-1.0ex]
			\textbf{Fraud specific} & \multicolumn{2}{l}{$\sbullet[.75]$ Excl. non-fraud specific literature} & \multicolumn{2}{c}{204} \\
			\hline \\[-1.0ex]
		\end{tabular}
	\end{adjustbox}
\end{table} 

\subsubsection*{Step 4 - Reading and Analysis:}

We read and analysed a total of 294 records. A total of 204 of these records are immediately related to a specific fraud analytical domain. The other 90 records, on the other hand, are domain-unspecific, i.e., propose solutions and methods that are not benchmarked on fraud-specific data or use cases. The entire selection of records can be found in the Appendix of this paper in Subsection ``Record Selection".

The remaining approximately 200 records are screened and evaluated along a well-defined set of dimensions, namely, a) their domain of application, b) the fraud analytical challenges faced as well as c) the performance metrics and d) fraud analytical methods used. To allow for cross-dimensional analysis, we structure our entire analysis in a relational database. The full design of the database can be found in Figure~\ref{fig:Database} in the Appendix. Additionally, an up-to-date version of the database including additions and revisions is available online and presented further in Section~\ref{sec:Database}. We elaborate on each of the dimensions we analyze and on their underlying motivation below:

\textit{Domains:} Domains are the specific applications of fraud analytical research and refer to the real-life use cases of methods and frameworks developed. \textcite{Phua.2010} provide an initial view on these, as do other existing reviews and surveys. We are especially interested in the most prominent domains from the last decade. This is an essential dimension to bridge from theory to practice. An overview of domains allows both researchers and practitioners to understand the focus of academic research and identify gaps between industry needs and academic efforts.

\textit{Challenges:} Challenges refer to the unique issues faced by researchers in developing data-driven fraud detection solutions. This effort aims to raise awareness of typical issues in fraud analytics, as well as to bridge between different domains and possibly even unrelated research fields.

A thorough inspection of the challenges mentioned and approached in research can further bridge the gap between theory and practice. We identified challenges by systematic screening of records. We do not restrict ourselves to challenges explicitly mentioned by the authors, for example, in the contributions, but carefully examine the whole paper for any demands, issues and concerns mentioned.

\textit{Performance metrics:} We strive to understand which metrics are commonly used in fraud analytics. The data might allow us to judge implicit biases in the evaluation of data-driven methods. 

\textit{Methods:} Focusing on fraud analytics, data-driven methods are at the core of the field. We aim to elicit which methods are used most frequently in fraud analytical research. Additionally, we will perform a cross-dimensional analysis of the methods. In other words, we aim to understand which methods are used in which domains and are addressing which challenges in the literature. Hereby, we aim to explore how academics handle specific fraud detection challenges and also to guide practitioners in selecting methods that might be suited to their respective domain of application. Due to the vast amount of different data-driven methods presented and used in the literature, we propose a structuring framework for data-driven methods. The framework is discussed in Section~\ref{sec:Method_Framework}.

\subsubsection*{Step 5 - Synthesis:}

Finally, we analyse our data in a structured way and present the results in Section~\ref{sec:Results}.

\subsection{A Clustering Framework for Fraud Analytical Research\label{sec:Method_Framework}}

To organize the large amount of different data-driven methods presented in the fraud analytical literature, we introduce and adopt a novel clustering framework for data-driven methods in fraud detection. To our knowledge, such a well-defined and agreed-upon framework has not yet been proposed, although we believe that it will help fellow researchers and readers make sense of previous research and the possibilities presented therein. Our framework is created for data-driven methods in fraud detection and differentiates itself from other established data mining and knowledge discovery frameworks, for example, the CRISP-DM model \parencite[]{Shearer.2000}, by focusing less on the overall knowledge discovery process and more on the methods used within. Due to its generality, we believe that it can be adopted in other data analytical fields besides fraud analytics. The framework is presented in Figure~\ref{fig:Clustering} and further described below:

\begin{figure}[ht]
	\centering
	\footnotesize
	\includegraphics[width=0.85\textwidth]{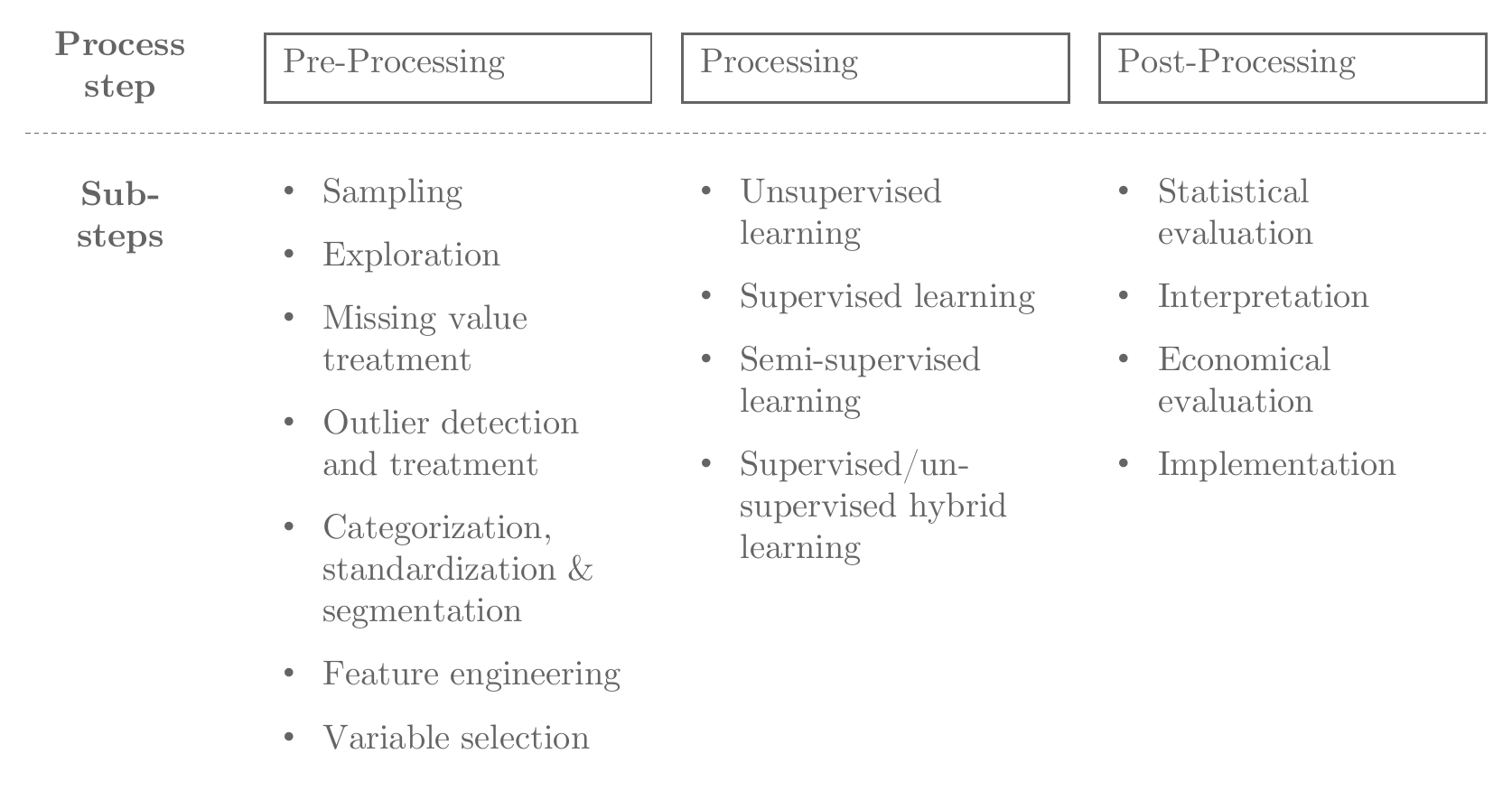}
	\caption[Method clustering]{Method clustering}
	\label{fig:Clustering}
\end{figure}

The framework is set up in two layers. The first layer concerns a categorization of methods into three high-level groups: pre-processing, processing and post-processing. This initial clustering is widely accepted in the literature and in practice \parencite[][]{Baesens.2014}. The second layer specifies subgroups of methods in each high-level category of the first layer.

\subsubsection*{Preprocessing:}

We define ``Preprocessing" as the step prior to training models and making prediction. More specifically, ``Preprocessing" deals with manipulating and cleaning data to make it suited for ``Processing"\footnote{Typically, this extends beyond enablement of ``processing" techniques to being able to use the data, but includes improvements in performance and interpretability of an analysis}. To further break this down, we adopt the preprocessing steps presented by \textcite{Baesens.2014}, i.e., ``Categorization, standardization \& segmentation", ``Exploration", ``Missing value treatment", ``Outlier detection and treatment", ``Sampling" and ``Variable selection". Additionally, we add ``Feature engineering" as a separate step, which is not mentioned by \textcite{Baesens.2014}, yet is necessary and frequently found in the literature (see Section~\ref{sec:Results}). ``Feature engineering" is especially relevant to date, due to the increasing presence of unstructured data in real-life fraud analytics, for example, textual or image data in claims handling at insurance companies.

\subsubsection*{Processing:}

We define ``Processing" as applying data-driven methods and training or learning data-driven models to data. ``Processing" is at the core of fraud analytics and is most frequently dealt with in research (cf. Section~\ref{sec:Results}). To obtain a collectively exhaustive framework, we decide to divide processing methods into supervised and unsupervised methods, as well as respective mixtures. This classification is an adoption of the clustering presented by \textcite{Phua.2010}. Their work differentiates between ``Supervised learning", ``Semi-supervised learning", ``Unsupervised learning" and ``Supervised/unsupervised hybrid learning".

\subsubsection*{Postprocessing:}

Compared to the other two steps, ``Postprocessing" is the least clearly defined step in the literature. We define ``Postprocessing" as all steps after ``Processing": for example, interpretation and evaluation. We introduce a novel clustering of postprocessing techniques. We divide ``Postprocessing" into four separate subcategories. ``Interpretation" is defined by all actions necessary to understand models and methods created and trained during ``Processing". ``Implementation" refer to techniques and steps necessary to operationalize predictions and outcomes of models and methods. ``Statistical evaluation" defines all steps necessary to evaluate the quality of a model or method from a pure statistical and research perspective. In contrast, ``Economical evaluation" defines steps necessary to evaluate a model or method in an applied setting, typically involving cost and time metrics. The presented clustering resembles the framework presented by \textcite{Bruha.2000}, yet we believe that it further adds tangibility and takes a step towards more closely resembling the postprocessing steps taken by practitioners and researchers.

\section{Results and Discussion\label{sec:Results}}
In this section we provide the key details from our review. All discussions are structured according to the dimensions presented in Section~\ref{sec:Method}.

\subsection{Paper Types and Periodicals\label{sec:Results_Types}}

Approximately two-thirds of the records in our database are published in academic journals, and only one-third are published in conference proceedings (cf. Figure~\ref{fig:Periodical} in the Appendix). By the nature of our sampling (cf. Section~\ref{sec:Method}), this is not representative of all fraud analytical publications. However, we conclude that the most prominent fraud analytics research is journal-based in the majority. Likewise, most records present methodologies or frameworks (approximately 80\%), whereas only one in five records in our database are surveys and reviews. Compared to the total population of fraud analytical literature, we likely face an overrepresentation of surveys and reviews, due to the high average citations of such papers \parencite[][]{Seglen.1997, Dong.2005, Simons.2008}.

\subsection{Domains\label{sec:Results_Domains}}

We identify 19 distinct fraud domains addressed in our literature sample. To limit the scope of this paper, we reduce this list to the five main and most prominent topics in fraud analytics, also presented in Figure~\ref{fig:Domains}. Additionally, we provide a distribution of domains across years in Figure~\ref{fig:Domains_over_time} in the Appendix. For a complete list of domains, we refer to our online database (cf. Section~\ref{sec:Database}). Below, we shed light on the five most important domains according to our study:

\begin{figure}[ht]
	\centering
	\footnotesize
	\includegraphics[width=0.85\textwidth]{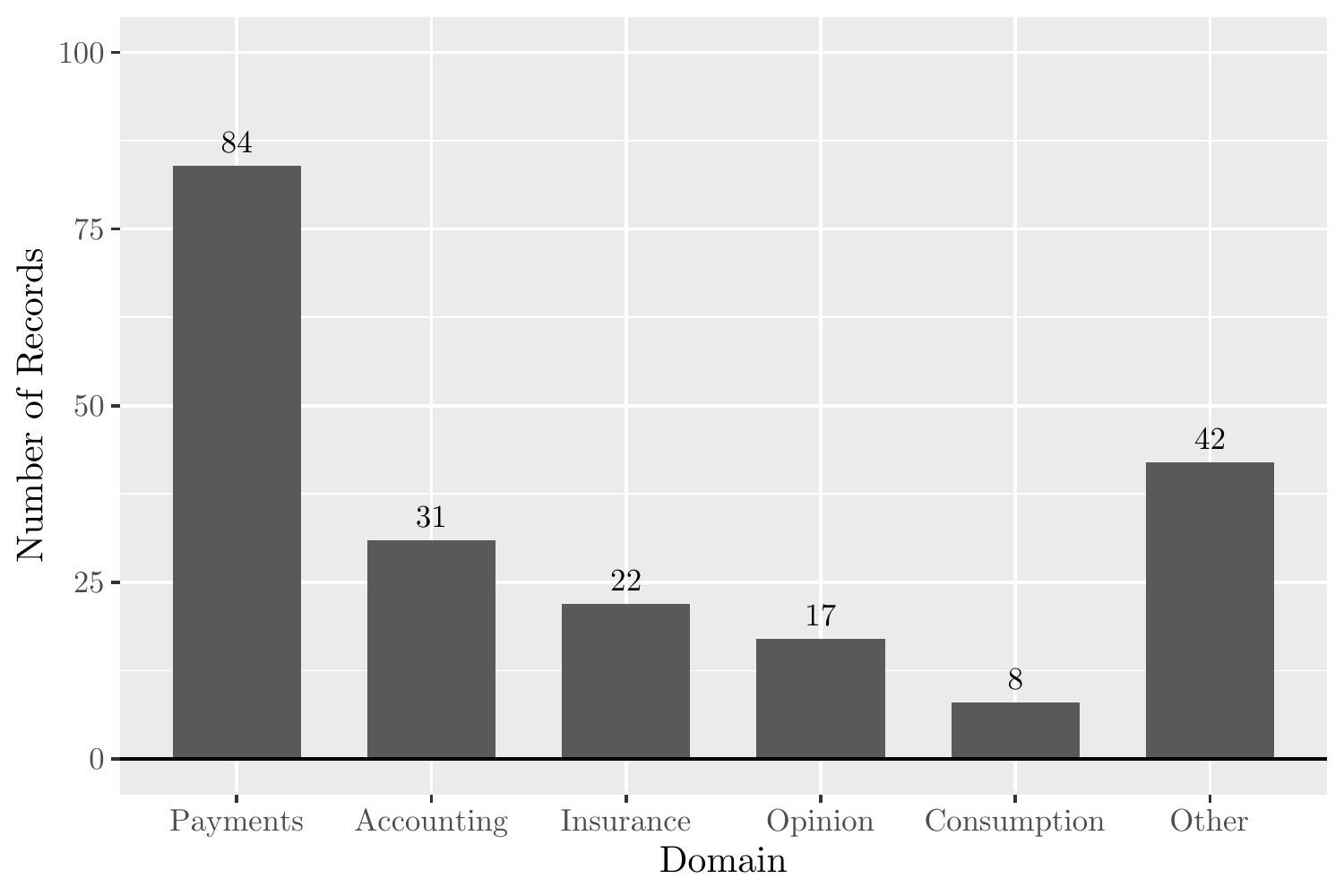}
	\caption[Distribution of domains]{Distribution of domains}
	\label{fig:Domains}
\end{figure}

\subsubsection{Payment Fraud.\label{sec:Results_Domains_Payments}}

Eighty-four or approximately 40\% of records pertain to with payment fraud. A majority of records classified in this domain deal with ``credit card fraud", wich is likely the most prominent specific use case in fraud analytics. However, we consider a variety of papers not explicitly dealing with credit card payments, but, for example, online payments \parencite[][]{Carminati.2015, Wei.2013} or debit card payments \parencite[][]{Nami.2018}. To leverage the similarity between those applications, both in data and business, we decided to combine them in a joint domain. Payment fraud can be committed in both ``business to customer (B2C)" and ``business to business (B2B)" settings. Based on our analysis of the literature, we define payments fraud as follows: 

\begin{definition}
\label{def:payments_fraud}
\textit{Payment fraud is the unlawful act of forging, stealing or manipulating personal or financial information to make payments on behalf of someone else, or to prevent real monetary transactions between the recipient of a good or service and the provider.}
\end{definition}

\subsubsection{Corporate and Accounting Fraud.\label{sec:Results_Domains_Corporate}}

Thirty-one or approximately 15\% of records deal with corporate and accounting fraud. The field addresses all types of fraud that are committed by corporations and businesses and that are linked to or pertaining to their finances. These fraud types include money laundering, accounting fraud and tax evasion. We define the domain as follows:

\begin{definition}
\label{def:accounting_fraud}
\textit{Corporate and accounting fraud concerns all fraud types committed by (representatives of) corporations to alter their finances. This includes the attempts to achieve monetary gain and to deceive of stakeholders.}
\end{definition}

The focus of the domain on corporate frauds of a financial nature is to reduce ambiguity and overlap with other fraud domains.

\subsubsection{Insurance Fraud.\label{sec:Results_Domains_Insurance}}

Insurance fraud relates to all actions aiming to deceive an insurer, typically meant to unlawfully claim (higher) insurance benefits or lower premiums. This type of fraud is comparable to ``payments fraud" (cf. Section~\ref{sec:Results_Domains_Payments}) and can happen in both B2B and B2C environments. The literature, however, predominantly addresses B2B applications for insurance fraud. Records on insurance fraud account for approximately 10\% of the total records in our database. The largest field within insurance fraud is health care insurance fraud. 

We define insurance fraud as follows:

\begin{definition}
\label{def:insurance_fraud}
\textit{Insurance fraud describes all actions of an insurance client to deceive the insurer and unlawfully claim (higher) benefits or to benefit from a lower premium (for example, by information misrepresentation).}
\end{definition}

\subsubsection{Opinion Fraud.\label{sec:Results_Domains_Opinion}}

Opinion fraud describes actions to manipulate either the opinion of others or reception of an opinion. Research focuses mostly on internet opinion fraud and more specific use cases include ``spam detection" and ``review fraud" on social networks and e-commerce. This fraud type accounts for approximately 10\% of records in our database. We define:

\begin{definition}
\label{def:opinion_fraud}
\textit{Opinion fraud is the act of trying to alter the opinion of an individual or group of individuals by changing available information or providing wrong information.}
\end{definition}

\subsubsection{Consumption Fraud.\label{sec:Results_Domains_Consumption}}

The final fraud type we describe is consumption fraud. This domain accounts for approximately 5\% of the total records in our literature sample. In our sample, all records related to consumption fraud pertain to identifying nontechnical energy loss, that is finding illegitimate consumers in electricity grids. We extend this specific domain to 'consumption', including, for example, telecommunications. This step aids pertaining to general and widely applicable definitions of fraud types. Explicitly, we describe:

\begin{definition}
\label{def:consumption_fraud}
\textit{Consumption fraud is the action of illegitimately consuming goods or taking services, by hiding either parts or the entire consumption from the vendor.}
\end{definition}

\subsection{Challenges\label{sec:Results_Challenges}}

\begin{figure}[ht]
	\centering
	\footnotesize
	\includegraphics[width=0.85\textwidth]{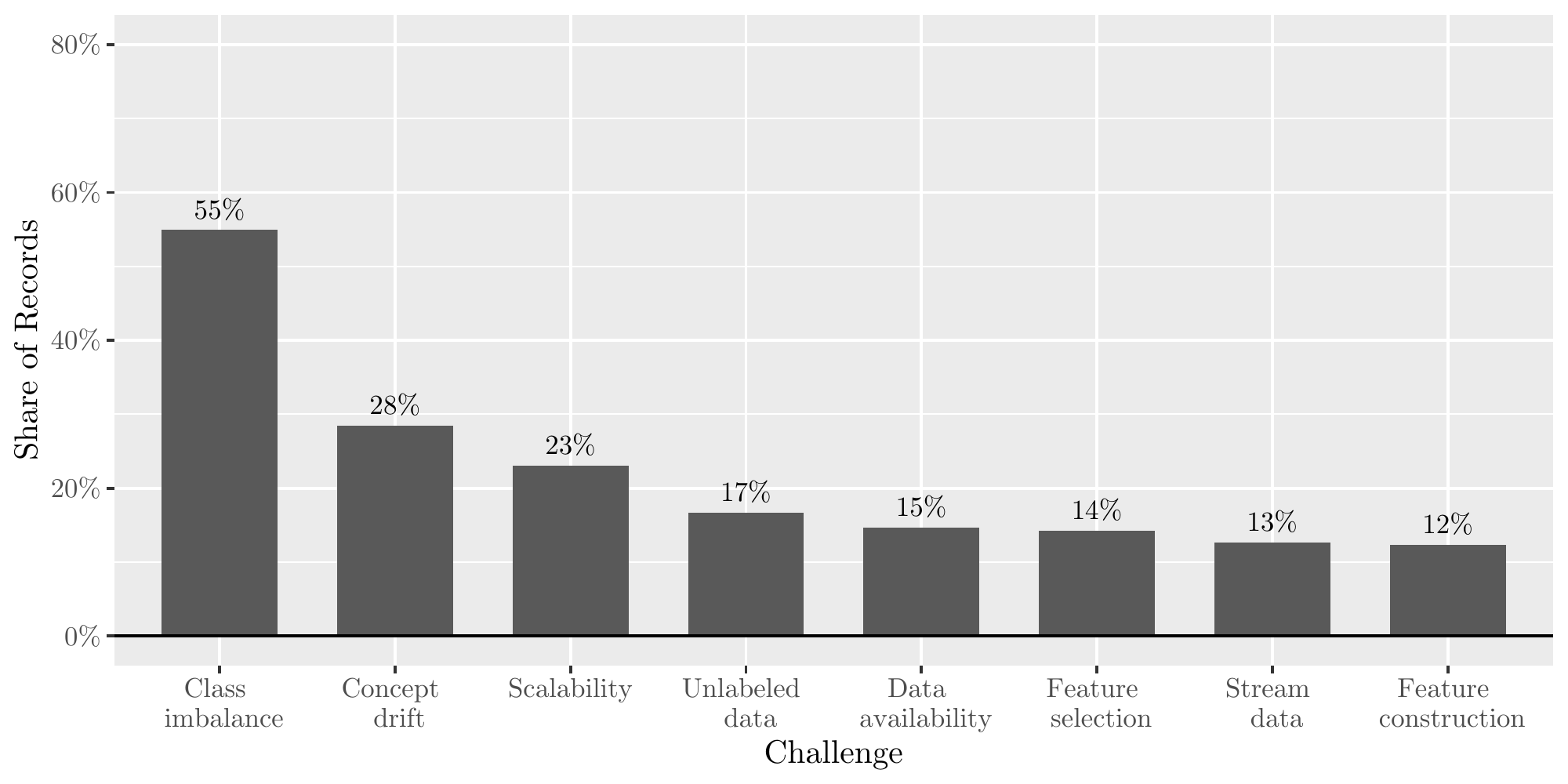}
	\caption[Key challenges]{Key challenges}
	\label{fig:Challenges}
\end{figure}

We identify 23 individual challenges in the fraud analytical literature. A full list and description of these can be found in Table~\ref{tbl:Challenges}. Furthermore, we will elaborate on the top eight most mentioned challenges. Figure~\ref{fig:Challenges} shows the share of records mentioning or dealing with these challenges\footnote{Note that challenges are issues discussed and reflected by researchers: they are not general tasks in processing fraud analytical data. Challenges need to be directly or indirectly mentioned as such by the authors in order to be considered in our analysis.}. Below, we will discuss each of these challenges in detail:

\begin{table}[p] 
	\caption[Fraud analytical challenges]{Fraud analytical challenges}
	\label{tbl:Challenges}
	\centering
	\begin{adjustbox}{max width=15.2cm}
		\renewcommand{\arraystretch}{0.8}
		\begin{tabular}{@{\extracolsep{2pt}}lp{12cm}} 
			\\[-1.0ex]\hline 
			\hline \\[-1.0ex] 
			{Challenge} & Description 	\\\hline \\[-1.0ex]			
			Automation & Build models that require a low degree of maintenance and human intervention along all dimensions of the data discovery process.\\[0.1ex]
			\hdashline \\ [-1.0ex]
			Categorical data & Enable analytical methods to deal with and utilize categorical data.\\[0.1ex]
			\hdashline \\ [-1.0ex]
			Class imbalance & Classes in the data are heavily unbalanced. One or more classes are in a minority against a majority class, impacting the performance of analytical methods.\\[0.1ex]
			\hdashline \\ [-1.0ex]
			Class rarity & Specific classes only appear rarely, likely not exhaustively representing the underlying class concept.\\[0.1ex]
			\hdashline \\ [-1.0ex]
			Concept drift & The underlying distribution of the data, or its concept, is time variant.\\[0.1ex]
			\hdashline \\ [-1.0ex]
			Data availability & Data for training and evaluating models is sparsely available or not available at all.\\[0.1ex]
			\hdashline \\ [-1.0ex]
			Feature construction & Raw data needs to be preprocessed and features need to be created for model training.\\[0.1ex]
			\hdashline \\ [-1.0ex]
			Feature selection & Data is high-dimensional and feature selection must be done before training.\\[0.1ex]
			\hdashline \\ [-1.0ex]
			Imbalanced instance costs & Costs of (mis-)classification vary between classes and instances.\\[0.1ex]
			\hdashline \\ [-1.0ex]
			Interpretability & Analytical methods require a high degree interpretability and explainability.\\[0.1ex]
			\hdashline \\ [-1.0ex]
			Method selection & A large body of analytical methods is available for analysing and detecting fraud with no clear indications to choose one over the other.\\[0.1ex]
			\hdashline \\ [-1.0ex]
			Mislabeled data & Training and test data are (partially) mislabeled.\\[0.1ex]
			\hdashline \\ [-1.0ex]
			Network data & Training and test data are structured in networks and graphs, as opposed to tabular data.\\[0.1ex]
			\hdashline \\ [-1.0ex]
			Noisy data & Training and test data include a high degree of noise and classes are not fully separable.\\[0.1ex]
			\hdashline \\ [-1.0ex]
			Stream data & Models need to learn and work on-the-go with new data entering live and in-stream.\\[0.1ex]
			\hdashline \\ [-1.0ex]
			Overfitting & Models overfit training data.\\[0.1ex]
			\hdashline \\ [-1.0ex]
            Performance evaluation & Standard performance metrics for evaluating fraud analytical methods do not satisfy practitioner demands.\\[0.1ex]
			\hdashline \\ [-1.0ex]
			Real-time execution & Trained methods must be able to operate in real time and have minimal delay.\\[0.1ex]
			\hdashline \\ [-1.0ex]
			Scalability & Fraud analytical methods must work and be able to cope with increasing amounts of data.\\[0.1ex]
			\hdashline \\ [-1.0ex]
			Sequential classification & Data points in training and test data are dependent and form time series.\\[0.1ex]
			\hdashline \\ [-1.0ex]
			Unlabeled data & Parts of the training data are missing labels completely.\\[0.1ex]
			\hdashline \\ [-1.0ex]
			Unstructured data & Training and test data are completely unstructured, e.g., text or image data.\\[0.1ex]
			\hdashline \\ [-1.0ex]
			Verification latency & Labels of training data are not immediately available to the fraud analytical methods, but are verified later in time.\\[0.1ex]
			\hline \\[-1.0ex]
		\end{tabular}
	\end{adjustbox}
\end{table} 

\subsubsection{Class Imbalance.\label{sec:Results_Challenges_classimb}}

By far, the most widely mentioned and addressed challenge in fraud analytical research is class imbalance. Over 50\% of records mention or refer to class imbalance as a challenge in research. ``Class imbalance" is present when class labels in a data set or data stream have contrasting prior probabilities \parencite[]{Japkowicz.2002}. The class with a high prior probability is called the ``majority class", and the class with a (very) low prior probability is called the ``minority class". This is the nature of many fraud analytical use cases: for example, in credit card fraud, a subdomain of ``payments fraud" (cf. Section~\ref{sec:Results_Domains_Payments}). In credit card payment streams, legitimate clients make up the vast majority of total payments. Fraudsters only contribute a small amount and share of payments. For illustration, a popular credit card fraud data set presented by \textcite{Pozzolo.2015}\footnote{A copy of the data set is available via \texttt{https://www.kaggle.com/mlg-ulb/creditcardfraud}} and includes only 492 fraudulent transactions (0.172\%) among a total of 284,807 transactions. 

According to our analysis of the literature, two problems emerge from class imbalance. One problem is that several data-driven methods require balanced classes for learning in order to not overfit the majority class: for example, decision trees and logistic regression. Second, when the prior probability of the minority class is sufficiently low and the training data set small, the instances of the minority class become ``rare": that is, there is not enough data to accurately learn the concept of the minority class \parencite[]{Bauder.2018b}. The latter is a problem often overlooked in the literature when dealing with class imbalance.

\subsubsection{Concept Drift.\label{sec:Results_Challenges_Conceptdrift}}

The second most noticed challenge is concept drift. Concept drift occurs when the underlying distribution, the concept, of a class within a data set is time dependent. As class imbalance, the topic is widely discussed in the general academic literature \parencite{Webb.2016, Wang.2016, Lu.2018} and is a problem in the nature of fraud. As investigators come to know and understand existing concepts of fraud, the rate of successful fraud would, ceteris paribus, decrease. Hence, fraudsters are required to adapt their approaches, that is, to change the concept. This is one of the reasons why, even though fraud analytical research has made significant progress over the years, fraud has not yet disappeared from most, if not all domains of application. An investigation of \textcite{UKFinance.2021}, for example, shows that losses from debit and credit card fraud have even increased in the last 10 years, ranging from approximately \textsterling 340 to \textsterling 670 million. However, it should be noted that, compared to the overall activity of fraudsters, this is still a success for fraud investigators \parencite[cf.][]{FTC.2020}.

In fraud analytics, concept drift typically introduces problems due to the nature of training data-driven models. When the concept of fraud changes over time, models will no longer be sensitive to new fraud cases and will no longer be able to detect them, resulting in high false negative rates and severe fraud damages when applied in real life.

\subsubsection{Scalability.\label{sec:Results_Challenges_scalab}}

The third challenge we detail is scalability, which is noticed in roughly one out of four records. Scalability, in itself, is a requirement for fraud analytical models to be able to efficiently and effectively handle large amounts of data. The challenge lies in satisfying this requirement. In most fraud analytical domains of application (cf. Section~\ref{sec:Results_Domains}), increasing digitization (e.g., in telecommunications) and a wider rate of adoption (e.g., in online payments) have not only increased the amount of agents creating data but also the amount of data created per capita \parencite[]{Hilbert.2011, Hilbert.2012}.

None of the selected papers dealt solely with scalability issues in fraud analytics, yet many authors dedicate attention and effort to benchmarks on scalability or proposing efficient methods. There are two notions of scalability. First, a method needs to be able to handle large data sets in the first place, both in training and application. Second, a method needs to maintain sufficient performance when handling large data sets: that is, it must be fast and maintain precision when scaling up the size of both training data and the amount of predictions in a real-life application.

\subsubsection{Unlabeled Data.\label{sec:Results_Challenges_unlabeleddata}}

Another challenge in fraud analytics is introduced by unlabeled training data. Based on our review of the literature, we distinguish between two cases: fully and partially unlabeled data. Whereas the fully unlabeled case is a suitable target for unsupervised learning techniques \parencite{Hastie.2001}, the partially unlabeled case adds another level of complexity. We found two separate cases of partially unlabeled training data. One case describes situations in which labels are simply not available for a subset of observations. The other case describes situations in which labels are given, yet in which the certainty that the label is correct is smaller than 100\%. 

This confronts researchers and practitioners with several problems: for example, fraud analytical methods must be robust to handle missing labels, and methods must be able to either handle uncertainty or detect mislabeled instances in the training data.

\subsubsection{Data Availability.\label{sec:Results_Challenges_dataavail}}

Data availability is another critical challenge in fraud analytical research. To build and test reliant and effective methods and models, researchers require representative samples of fraud data to analyze and learn from. However, many fraud analytical domains of application deal with private or semiprivate information, and real-life practitioners are confronted with confidentiality and privacy requirements, restricting them from sharing data with a wider audience. It is no surprise that, though an applied science, fraud analytics lacks a wide selection of real-life and up-to-date data sets and partially relies on synthetic data sets. The impact of outdated data on concept drift is described in Section~\ref{sec:Results_Challenges_Conceptdrift}.

This challenge introduces several problems, including working on unrepresentative data and working with outdated data. Hence, researchers risk proposing methodologies that are not tailored explicitly for real-life applications. In such a setting, the true concept might be different from the data used in research or may have drifted.

We are convinced that the impact of a lack of training data and the reliance on a small set of data sets is underresearched. For potential future research, we refer to Section~\ref{sec:ResearchDirs}.

\subsubsection{Feature Selection.\label{sec:Results_Challenges_featuresele}}

``Feature selection" refers to the challenge of selecting the right predictors from a large number of available variables for training methods and detecting fraud. The challenge is closely related to ``High dimensionality", that is, the case in which the total number of predictors outnumbers the number of available training observations \parencite[]{Buhlmann.2011}, and multicollinearity, wherein explanatory variables in the data set are correlated with each other \parencite[]{Hastie.2001}. Additionally, ``Feature selection" is often linked to issues of ``Interpretability" in finding variables that explain fraud well and simultaneously are interpretable for practitioners.

\subsubsection{Stream Data.\label{sec:Results_Challenges_streamdata}}

``Stream data" refers to the challenge of working with continuous flows of information, both for training and predicting observations. Depending on the domain of application and the specific setting, fraud analytical models might be required to learn ``online": that is, to update live based on every input observation. Naturally, this challenge is related to both scalability (cf. Section~\ref{sec:Results_Challenges_scalab}) and concept drift (cf. Section~\ref{sec:Results_Challenges_Conceptdrift}). Typically, in situations with relevant stream data occurring, the amount of information is large, and concept drift can appear rapidly.

The problem of online learning (also referred to as incremental learning) and working with stream data is widely discussed in the literature. For further reading, we refer to relevant records on the topic, including \textcite{Zhu.2010}, \textcite{He.2011} and \textcite{Krawczyk.2017}.

\subsubsection{Feature Construction.\label{sec:Results_Challenges_Featureconstr}}

``Feature construction" is another well-known challenge and aspect of machine learning and analytics in practice. The challenge refers to the ``preprocessing" step of the same name (cf. Section~\ref{sec:Method_Framework}). It is associated with quantitative and qualitative analysis of the data and modifying it to enable more efficient and effective learning. Feature engineering becomes especially important in situations with network information or graph structures between agents or observations, qualities that traditional processing techniques cannot deduct from, for example, tabular data.

For an elaborate overview of preprocessing strategies and techniques, we refer to the present body of literature, for example, \textcite{Dong.2018b} or \textcite{Zheng.2018c}.

\subsection{Methods\label{sec:Results_Methods}}

For the analysis of methods, we take into account all methodological papers from 2011 to 2018. This is due to the highly volatile quality of papers from recent years, a result of our citation-based search strategy\footnote{This approach results in at least three years of time which we consider sufficient to establish a meaningful number of citations across the papers we investigate. Note that we do not perform this filtering step for other dimensions, as both domains and challenges are more homogeneous than methodologies: that is, there are fewer.}. This approach reduces the number of papers relevant for this analysis to 94. 

87\% of records deal with methods targeting ``processing". Only 38\% and 16\% deal with either \mbox{``pre-"} or ``postprocessing". Most of the well-received papers present a methodology or framework new to fraud analytics, which is applied to a relatively clean and ready-to-use data set, yielding this distribution.

\begin{figure}[ht]
	\centering
	\footnotesize
	\includegraphics[width=0.95\textwidth]{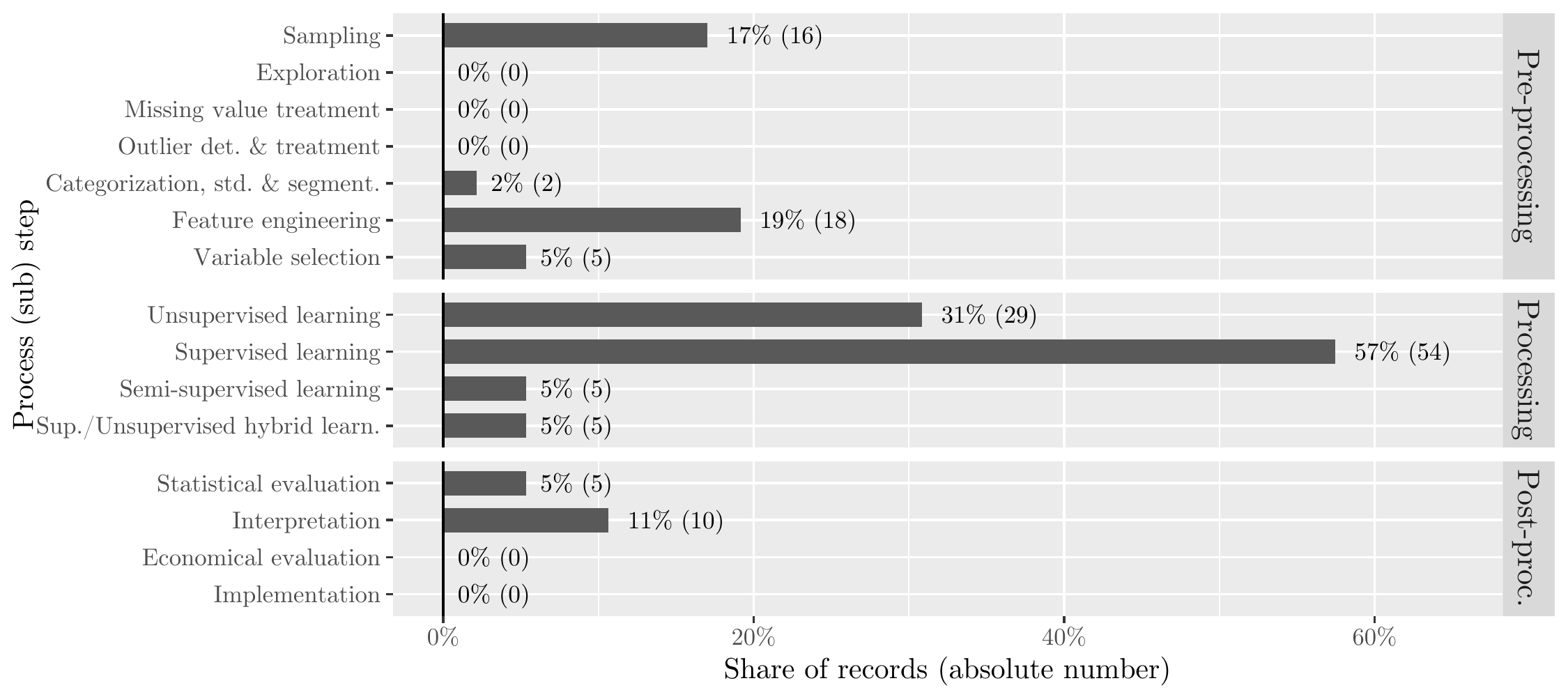}
	\caption[Processing steps by share of records]{Processing steps by share of records}
	\label{fig:Framework_steps_analysis}
\end{figure}

The granularity of this analysis is set by applying the classification framework for fraud analytical methods proposed in Section~\ref{sec:Method_Framework}. Our framework completely covers the analytical steps available in fraud analytics. Investigating which steps are commonly discussed in the literature enables us to accurately understand research efforts. The share of records covering each of the process steps and process substeps is depicted in Figure~\ref{fig:Framework_steps_analysis}.

In ``Preprocessing", we find the largest interest in the literature in substeps ``Feature Engineering" and ``Sampling". Evidently, this is in line with the associated challenges we have discussed in Section~\ref{sec:Results_Challenges}. Feature engineering itself is both a challenge and a step necessary to be taken in fraud analytics. Sampling, on the contrary, a step crucial for many data-driven methods to counteract problems associated with imbalanced class distributions. We find no record that actively discusses applying techniques in ``Exploration", ``Missing value treatment" or ``Outlier detection and treatment". We see several reasons for this. First, ``Exploration" is related to both data and domain understanding. This is often considered redundant, as records actively position themselves in fraud analytics and do not engage in narrowly defining what either fraud or their specific domain of application is (this is, after all, a reason why we decided to write this work). Additionally, the limited availability of data sets hinders researchers from diving deeper into exploration by forcing them to work on the available data and stories there. We refer to the popular ``CRISP-DM" framework for data mining \parencite[]{Shearer.2000} to further illustrate our point. ``CRISP-DM" does include a ``data understanding" step, yet this is closely entangled with ``business understanding", a step that, for many researchers, is not possible to conduct.

\begin{figure}[ht]
	\centering
	\footnotesize
	\includegraphics[width=0.75\textwidth]{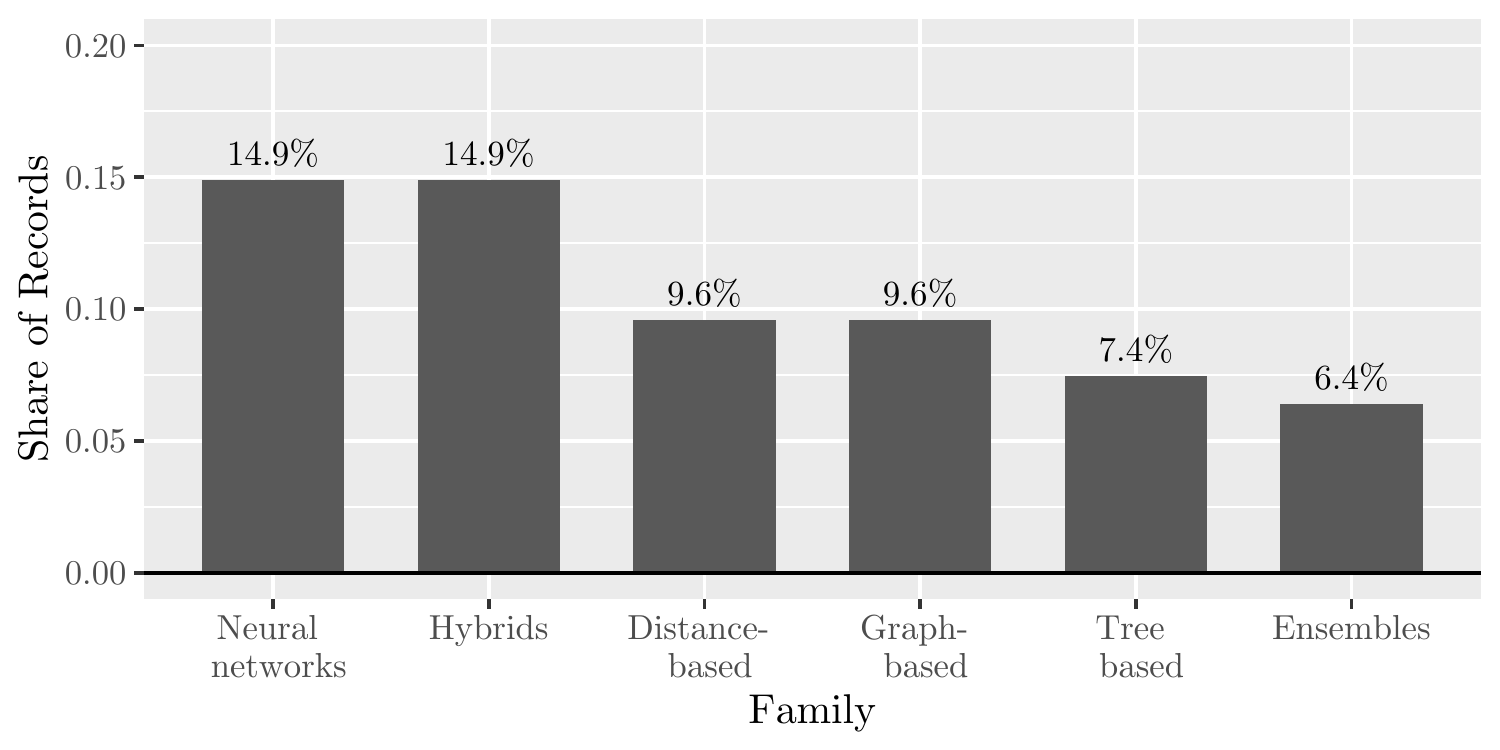}
	\caption[Families of processing techniques by share of records]{Families of processing techniques by share of records}
	\label{fig:Method_families}
\end{figure}

In ``Processing", the most popular type of technique is ``Supervised learning", referring to methods that learn from labeled training data. Approximately half of all records we analyzed dealt with supervised learning techniques. The second most popular type after ``supervised learning" is ``unsupervised learning", which is representative of all methods that do not require any labeled training data to predict fraud. ``Semi-supervised learning" techniques and ``Hybrid learning" are only relevant to a mere 5\% of records in our sample. In light of the limited amount of training data available to researchers in fraud analytics, this is a surprise. Many authors market unsupervised and semisupervised techniques as especially suited for situations where data are either sparsely available or of low quality, problems that are widely accepted and discussed in the general literature (cf. Section~\ref{sec:Results_Challenges}). Due to the popularity of ``Processing" in fraud analytics, we include a more detailed analysis of the records in our literature sample. Figure~\ref{fig:Method_families} shows the appearance of individual families of ``Processing" techniques in the 94 records we analyzed. We note the high popularity of perceptron-based (or neural network) techniques, as well as hybrid techniques, which are techniques that actively combine different families of methods. For the full data on all families of techniques, specific techniques per record and likewise data for preprocessing and postprocessing, we refer to our full results (cf. Section~\ref{sec:Database}).

Finally, in ``Postprocessing", the records in our sample solely pertain to either ``Interpretation" or ``Statistical evaluation". ``Implementation" and ``Economical evaluation" are not explicitly addressed. We believe that this is due to the same reasons that ``Exploration" is not prominently addressed in ``Preprocessing" (see above). The relative popularity of ``Interpretation", on the contrary, is likely a coproduct of the call for explainability in fraud analytics, another challenge we have distilled from the literature (cf. Figure~\ref{fig:Challenges}).

\subsection{Metrics\label{sec:Results_Metrics}}

\begin{figure}[ht]
	\centering
	\footnotesize
	\includegraphics[width=0.85\textwidth]{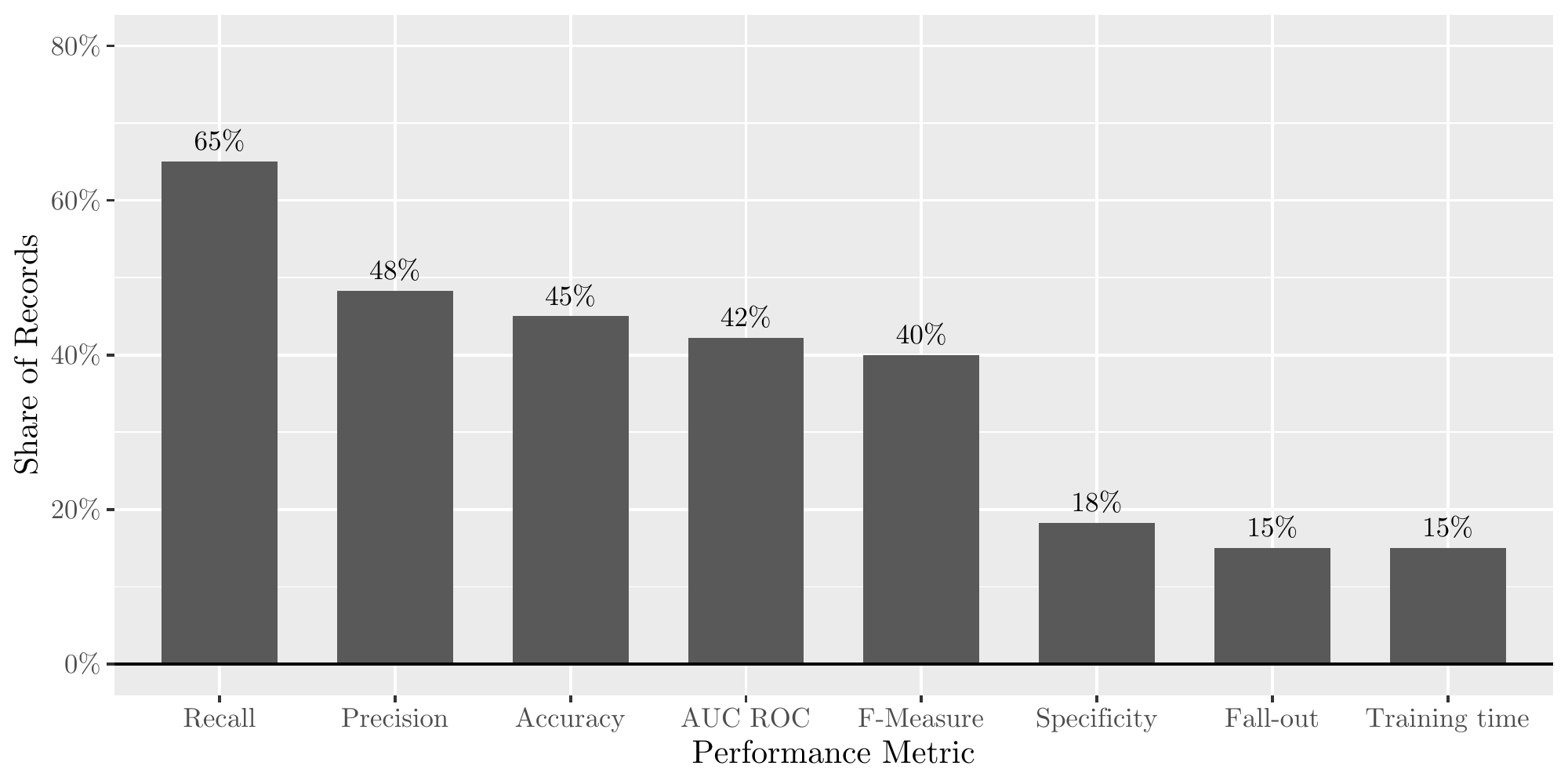}
	\caption[Commonness of performance metrics]{Commonness of performance metrics}
	\label{fig:Perf_Metrics}
\end{figure}

\noindent In this section, we present and discuss the most widely stated performance metrics in our literature sample. Figure~\ref{fig:Perf_Metrics} shows the top eight metrics by share of records. 

The most prominent metric we find is ``Recall". Recall intuitively describes which share of the fraudulent cases in a data set have been found by a method.

The second most stated performance metric is ``Precision", which describes the share of observations identified as fraudulent which are truly fraudulent.

``Precision" and ``Recall" are widely used performance metrics for classifying data with class imbalance. Hence, their popularity in fraud analytics is not surprising. Analogously, we observe wide use of the $F$-measure (also $F$-score), which is built upon precision and recall. All of the previously mentioned measures target overcoming the pitfalls of ``Accuracy" in environments with class imbalance. As highlighted in Section~\ref{sec:Results_Challenges_classimb}, in fraud analytics, researchers and practitioners are especially seeking models sensitive to identifying the minority class (fraudulent cases). A high accuracy, however, is only sensitive to a good fit of a model to the majority class.

In contrast, ``Accuracy" is still a widely stated performance metric and is found in almost every second record we analyzed. However, ``Accuracy" is usually accompanied by other metrics.

The fourth most stated performance metric in our literature sample is the ``Area under the Receiver Operating Characteristic Curve" (AUC ROC). It evaluates the trade-off between ``Sensitivity" (the ``True Positive Rate") and ``Specificity" (the ``True Negative Rate") of a method. AUC ROC is the only performance metric in the top eight that is threshold-invariant: that is, not dependent on a cut-off value.

One particularly striking performance metric in the top eight is ``Training time". This metric does not focus on the statistical performance of a method but solely investigates its computational complexity and the time needed to process data. In light of the challenge of ``Scalability", we are pleased to see this emerging in fraud analytics. ``Training time", however, is less clearly defined than traditional metrics and is dependent on a multitude of factors, for example, the size of the training data set, dimensionality of the data set or the learning setting (e.g., supervised or unsupervised learning).

Noteworthy is the low popularity of cost-based performance metrics and the minimal notion of economical evaluation of methods. We will elaborate further in Section~\ref{sec:ResearchDirs} on future research directions in fraud analytics.

Additionally, we note that the most popular performance metric, except AUC, is threshold based: that is, that a certain classification threshold has to be chosen to calculate the metric. The selection of this threshold is of key importance for the overall performance of a method and the real-world implications after implementation. We will again discuss this further in Section~\ref{sec:ResearchDirs}.

For an elaborate presentation and discussion of performance metrics in analytics, we refer the reader to the existing literature, for example, \textcite{Olson.2008}.

\subsection{Cross-Dimensional Results\label{sec:Results_CrossDim}}
Finally, we aim to bridge the different dimensions we have presented so far. We are, for the most part, interested in the cross-section of challenges and the respective methods used to solve them. This question is motivated by the challenge of ``Method selection" observed in the literature (cf. Table~\ref{tbl:Challenges}).

We find that many authors do not motivate the use of a method to solve a specific challenge in fraud analytics. We analyzed 94 records of data-driven methods (cf. Section~\ref{sec:Results_Methods}), and we identified 117 methods that were indicated by the authors as solutions to specific challenges. In comparison, we identified 276 methods in total. 

For our work, we will focus on the eight challenges that have been addressed by most methods in our sample of the literature. We state the number of methods dedicated to solving the respective challenge, both in total and per processing substep of our classification framework of fraud analytical methods (cf. Section~\ref{sec:Method_Framework}). The results are presented in Table~\ref{tbl:Methods_for_Challenges}.

\begin{table}[htbp] 
	\caption[Methods presented to solve fraud analytical challenges]{Methods presented to solve fraud analytical challenges}
	\label{tbl:Methods_for_Challenges}
	\centering
	\begin{adjustbox}{max width=15.2cm}
		\renewcommand{\arraystretch}{0.8}
		\begin{tabular}{@{\extracolsep{2pt}}lclc} 
			\\[-1.0ex]\hline 
			\hline \\[-1.0ex] 
			             & Methods  & Sub-step & Number of \\
			Challenge    & proposed & proposed & records \\
			\hline \\[-1.0ex]	
            Class imbalance & 29 & Sampling & 9 \\
             &  & Supervised learning & 9 \\
             &  & Sup./Unsupervised hybrid learn. & 4 \\
             &  & Semi-supervised learning & 3 \\
             &  & Unsupervised learning & 3 \\
             &  & Feature engineering & 1\\[0.1ex]
			\hdashline \\ [-1.0ex]
            Scalability & 15 & Supervised learning & 6 \\
             &  & Unsupervised learning & 4 \\
             &  & Semi-supervised learning & 3 \\
             &  & Sampling & 2\\[0.1ex]
			\hdashline \\ [-1.0ex]
            Feature engineering & 13 & Supervised learning & 5 \\
             &  & Feature engineering & 4 \\
             &  & Variable selection & 2 \\
             &  & Categorization & 1 \\
             &  & Semi-supervised learning & 1\\[0.1ex]
			\hdashline \\ [-1.0ex]
            Unlabeled data & 12 & Unsupervised learning & 9 \\
             &  & Feature engineering & 2 \\
             &  & Semi-supervised learning & 1\\[0.1ex]
			\hdashline \\ [-1.0ex]
            Interpretability & 10 & Interpretation & 8 \\
             &  & Semi-supervised learning & 1 \\
             &  & Unsupervised learning & 1\\[0.1ex]
			\hdashline \\ [-1.0ex]
            Concept drift & 9 & Supervised learning & 2 \\
             &  & Sup./Unsupervised hybrid learn. & 2 \\
             &  & Unsupervised learning & 2 \\
             &  & Feature engineering & 1 \\
             &  & Sampling & 1 \\
             &  & Semi-supervised learning & 1\\[0.1ex]
			\hdashline \\ [-1.0ex]
            Imbalanced instance & 8 & Supervised learning & 5 \\
            costs & & Statistical evaluation & 2 \\
             &  & Sampling & 1\\[0.1ex]
			\hdashline \\ [-1.0ex]
            Sequential classification & 6 & Supervised learning & 3 \\
             &  & Feature engineering & 1 \\
             &  & Sampling & 1 \\
             &  & Unsupervised learning & 1\\[0.1ex]
			\hdashline \\ [-1.0ex]
            Verification latency & 4 & Supervised learning & 2 \\
             &  & Semi-supervised learning & 1 \\
             &  & Sup./Unsupervised hybrid learn. & 1\\[0.1ex]
			\hline \\[-1.0ex]
		\end{tabular}
	\end{adjustbox}
\end{table} 

As with challenges overall, ``Class imbalance" again is the challenge that is addressed most by specific methods. In total, authors target the challenge 29 times by a specific method. The methods primarily used to target ``Class imbalance" are sampling and supervised learning techniques. The proposed solutions to ``Scalability" are typically found in processing techniques, foremost among them being ``supervised learning". An interesting notion is observed in tackling ``feature engineering". While a valid and established preprocessing step, the challenge is often targeted by processing techniques. Indeed, approximately half of the identified methods are situated in either ``supervised learning" or ``semi-supervised learning". A closer analysis of and reflection upon our cross-sectional results are left open for the reader.

For further analyses, we want to inspire fellow researchers to use our data or build upon it.

\section{Research Directions\label{sec:ResearchDirs}}

We highlight four directions for future research that we find critical for the research field, as well as to bridge from theory to practice. Additionally, we leave it open to fellow researchers to conduct further analyses of our work to detect gaps in their own domains and specializations. For additional information and data, see Section~\ref{sec:Database}.

\subsubsection*{Data Collection and Sharing.}

Further contributing to data collection and data sharing should be an imperative for all researchers in fraud analytics. Revisiting Definition \ref{def:fraud_analytics} of fraud analytics, researchers rely on the public availability of real-life data. With little data available and an increasing dependence on synthetic or old data sets (cf. Section~\ref{sec:Results_Challenges_dataavail}), fraud analytical research could face several risks. We see two ways to counter this development: 

On the one hand, in domains where data privacy is not a concern, further data collection in collaboration with industry practitioners will bring substantial value to research and the academic community. A promising direction, for example, would be the compilation of benchmarking data sets that both originate from different domains of application and that exhibit distinct fraud analytical challenges. To our knowledge, such a collection is not available to date. This would largely add to the positive impact of research by aligning research efforts and improving reproducibility. Practitioners could leverage the data as well, though this academic effort should not reduce the importance of interorganizational sharing of data \parencite[]{Barrett.1982}.

However, in domains where data privacy is an issue, research in cryptography would enable practitioners and researchers to share data in the first place. \textcite{Park.2022} provide a framework on how economic actors could share fraud concepts without leakage of private and critical information and provide the first relevant directions in consumption fraud detection\footnote{\textcite{Park.2022} explicitly focus on customs fraud, a subdomain that we did not encounter in our own sample of the literature.}.

With respect to the challenges in fraud analytics (cf. Section~\ref{sec:Results_Challenges} and Table~\ref{tbl:Challenges}), we define 8 requirements that data in fraud analytics should satisfy to be used in research. We present these requirements in Table~\ref{tbl:data_reqs}.

\begin{table}[ht] 
	\caption[Requirements for data sets in fraud analytics]{Requirements for data sets in fraud analytics}
	\label{tbl:data_reqs}
	\centering
	\normalsize
		\renewcommand{\arraystretch}{0.8}
		\begin{tabular}{@{\extracolsep{2pt}}>{\raggedright\arraybackslash}p{5cm}p{9cm}} 
			\\[-1.0ex]\hline 
			\hline \\[-1.0ex] 
			Requirement & Description\\
			\hline \\[-1.0ex]
			Context information & Context information is needed to understand the specific practitioner needs or business context of the data set\\[0.1ex]
			\hdashline \\ [-1.0ex]
			Interpretability of variables & For effective feature engineering and understanding of models, variables must be interpretable: hence, understandable variable names or a variable description is needed.\\[0.1ex]
			\hdashline \\ [-1.0ex]
			Label verification information & With uncertainty in label information and verification latency, data must represent either the certainty/uncertainty of labels or the time when the label was verified.\\[0.1ex]
			\hdashline \\ [-1.0ex]
			Misclassification costs per class/observation & Respective of the business context, misclassification costs should be made available either per class or (preferably) per observation to appropriately evaluate the economic implications of the model.\\[0.1ex]
			\hdashline \\ [-1.0ex]
			Preserved categorical information & Categorical information must be preserved to maintain a maximum amount of information in the data set.\\[0.1ex]
			\hdashline \\ [-1.0ex]
			Preserved time information & Each observation should include a timestamp to enable judging and treating concept drift.\\[0.1ex]
			\hdashline \\ [-1.0ex]
			Representative sampling time frame & To reflect concept drift (cf. Section~\ref{sec:Results_Challenges_Conceptdrift}), data must be sampled from a sufficiently large time frame.\\[0.1ex]
			\hdashline \\ [-1.0ex]
			Representative size & To check for scalability (cf. Section~\ref{sec:Results_Challenges_scalab}), the data set must be sufficiently large.\\
			\hline \\[-1.0ex]
		\end{tabular}
\end{table}

We leave it open for future research to test and evaluate existing fraud data sets for these qualities. As mentioned in Section~\ref{sec:Results_Challenges_dataavail}, attention should be paid to the impact of relying on a small number of data sets.

\subsubsection*{Establishment of Unified Performance Metrics.}

We hope to stimulate further advancements in the evaluation of fraud analytical methods and performance metrics. From our analysis of the literature, we find that the majority of records use global performance metrics, that is, metrics that evaluate based upon the entire data (cf. Section~\ref{sec:Results_Metrics}). Likewise, many of these metrics are threshold-based in nature, for example, sensitivity and specificity. Contrasting the notion that there is no wrong or right performance metric \parencite[]{Seliya.2009}, in a recent paper, \textcite{Carcillo.2021} in cooperation with an industrial partner state the importance of local performance indicators, such as the ``top-\textit{n} precision", a metric that evaluates the performance of a model only on the \textit{n} observations with the highest predicted confidence of being fraudulent. According to them, few practitioners have the capacity to investigate all potential fraud cases but need to efficiently allocate resources for further investigation. Similarly, the evaluation of methods based on induced costs focuses more on costly observations rather than the less costly (less important) majority of observations. While we do see some notion of cost-awareness in the literature, cost-sensitive learning and particularly cost-centric evaluation in fraud analytical research is still in the minority.

Additionally, with growing amounts of data and the need for more scalable methods (cf. Sections \ref{sec:Results_Challenges_scalab} and \ref{sec:Results_Metrics}), benchmarking existing and future models based on training time and scalability will be a key aspect of fraud analytics. This is in line with the performance metric ``training time", which we have found.

Future research on performance metrics and evaluation would contribute to further bringing together research and practice. Projects should extend from surveying actual industry needs in specific domains to investigating the impacts of choosing certain performance metrics over others. We expect that future fraud analytical research will have to focus more on aspects such as the performance of methods in terms of top-\textit{n}, operating costs and training time.

\subsubsection*{Unified Benchmarking Experiments for Fraud Analytical Methods.}

Third, we encourage the research community to build towards an established set of benchmarking experiments, essentially combining our proposals for more data sharing and unified performance metrics. 

These experiments should facilitate a set of unified data sets, objectives and performance metrics. Data should come from various application domains. The approach could help to further streamline the literature and research on fraud analytics by adding transparency to the strengths and weaknesses of models, both outside and inside their dedicated domain of application. The experiments could be extended over time to reflect trends and developments in the field.

The approach would be novel, especially since there has so far been little work in fraud analytics across different domains of application (cf. Section~\ref{sec:LitReview}).

\subsubsection*{Fraud Detection Systems to Incorporate Unstructured Data Sources.}

Finally, we see a need for research on fraud detection systems that enable leveraging external and unstructured data on top of the predominantly tabular data currently studied in fraud analytical research. In domains such as insurance fraud, relying solely on tabular data is not representative of real-life applications. In contrast, practitioners are faced with an increasing amount of additional, predominantly unstructured data to run their business and prevent fraud. Examples herein include pictures of damaged property, written reports or behavioral data \parencite[]{Weinmann.2021}.

An important future research topic should be the design and investigation of system architectures and frameworks that enable practitioners to leverage unstructured data on top of the established tabular data.

\section{A Database and Taxonomy for Research on Fraud Analytics\label{sec:Database}}
Our full database is published online under \texttt{https://tinyurl.com/fraudliteraturedb}. The database represents a snapshot of the literature and is neither complete nor perfect, though assembled with utmost care from the authors' side. We encourage fellow researchers to review the database and add their papers to it. Details on how to engage are listed online.

In addition, we found that to work and search the fraud analytical literature more efficiently, a keywording convention is needed for future publications. We are convinced that the dimensions of the literature we have investigated (cf. Section~\ref{sec:Method}) provide solid guidelines for keywording works on fraud analytics. We propose four types of keywords per paper to comply with our convention:

\begin{enumerate}
	\item \textit{``Fraud analytics"}
	\item \text{[\texttt{fraud analytical domain}]} in alignment with or as extension of Section~\ref{sec:Results_Domains}
	\item \text{[\texttt{fraud analytical challenges}]} in alignment with Table~\ref{tbl:Challenges}
	\item \text{[\texttt{fraud analytical methods}]} grouped by the clustering framework from Section~\ref{sec:Method_Framework}
\end{enumerate}

\section{Conclusion\label{sec:Conclusion}}

We surveyed a sample of the most prominent fraud analytical literature in the last decade (2011 to 2020). Our results provide a holistic overview of the field, including domains of application, emerging challenges and the methods used to detect and fight fraud. Additionally, we provide a clustering framework for fraud analytical methods, breaking them down into ``preprocessing", ``processing" and ``postprocessing" techniques and subcategories therein (cf. Figure~\ref{fig:Clustering}).

Our research extends the work of \textcite{Phua.2010}, and we continue to add value by narrowly defining and organizing critical aspects of fraud analytical research, e.g., a definition of fraud analytics (cf. Definition \ref{def:fraud_analytics}), as well as fraud analytical domains of application (cf. Definitions \ref{def:payments_fraud} to \ref{def:consumption_fraud}). Additionally, our work is unique in providing a summary of challenges in the detection, prevention and reduction of fraud across domains of application.

We surveyed almost 300 records on or related to fraud analytics. To make the amount of information consumable, we decided to structure it in a relational database. The database is available publicly (cf. Section~\ref{sec:Database}).

We place special emphasis on the future of fraud analytical research. We identified four areas critical to the future success of the field. All of them are targeted to ensure efficiency and effectiveness in research and to bridge between industry and practitioners. Future research in fraud analytics should especially focus on acquiring new and representative data sets for the evaluation and benchmarking of data-driven methods.

We hope that our work is able to shed light on fraud analytics in a novel and meaningful way and that it serves as a stepping stone for current and future researchers in the field. In light of ever-increasing fraud activities \parencite[]{FTC.2020}, we expect the field to continue and potentially increase its current popularity. By reducing the current ambiguity in research domains and research objectives, we believe that our work will play a relevant part in organizing and structuring.

Our future research focus will be on benchmarking existing fraud analytical data sets in research and on investigating the bias introduced by relying on only a small selection of data (cf. Section~\ref{sec:ResearchDirs}).

\newpage
\printbibliography[keyword=major-sources]

\newpage

\setcounter{table}{0}
\renewcommand{\thetable}{A\arabic{table}}
\setcounter{figure}{0}
\renewcommand{\thefigure}{A\arabic{figure}}

\begin{appendix}
	\section{Tables}

    \begin{table}[ht] 
    	\caption[Search Queries]{Search Queries}
    	\label{tbl:SearchQueries}
    	\centering
    	\begin{adjustbox}{max width=15.2cm}
    		\renewcommand{\arraystretch}{0.8}
    		\begin{tabular}{@{\extracolsep{2pt}}lp{12cm}} 
    			\\[-1.0ex]\hline 
    			\hline \\[-1.0ex] 
    			{Database} & Query 	\\\hline \\[-1.0ex]			
    			\textbf{Scopus} & \texttt{TITLE-ABS-KEY(("fraud detection"  OR  "fraud analytics") AND data) AND DOCTYPE(ar OR cp) AND (PUBYEAR > 2010 AND PUBYEAR < 2021) AND LANGUAGE(english) AND SUBJAREA(comp OR engi OR math OR busi OR deci OR econ)}\\[0.1ex]
    			\hdashline \\ [-1.0ex]
    			\textbf{Web of Science} & \texttt{(((TS=(("fraud detection" OR "fraud analytics") AND data) OR AB=(("fraud detection" OR "fraud analytics") AND data) OR TI=(("fraud detection" OR "fraud analytics") AND data)) AND PY=(2011-2020)) AND DT=(Article OR Proceedings Paper)) AND LA=(English) and Computer Science or Engineering or Business Economics or Science Technology Other Topics or Mathematics (Research Areas)}\\[0.1ex]
    			\hline \\[-1.0ex]
    		\end{tabular}
    	\end{adjustbox}
    \end{table} 
    
    \newpage
    \section{Figures}
    
    \begin{figure}[htbp]
    	\centering
    	\footnotesize
    	\includegraphics[width=0.75\textwidth]{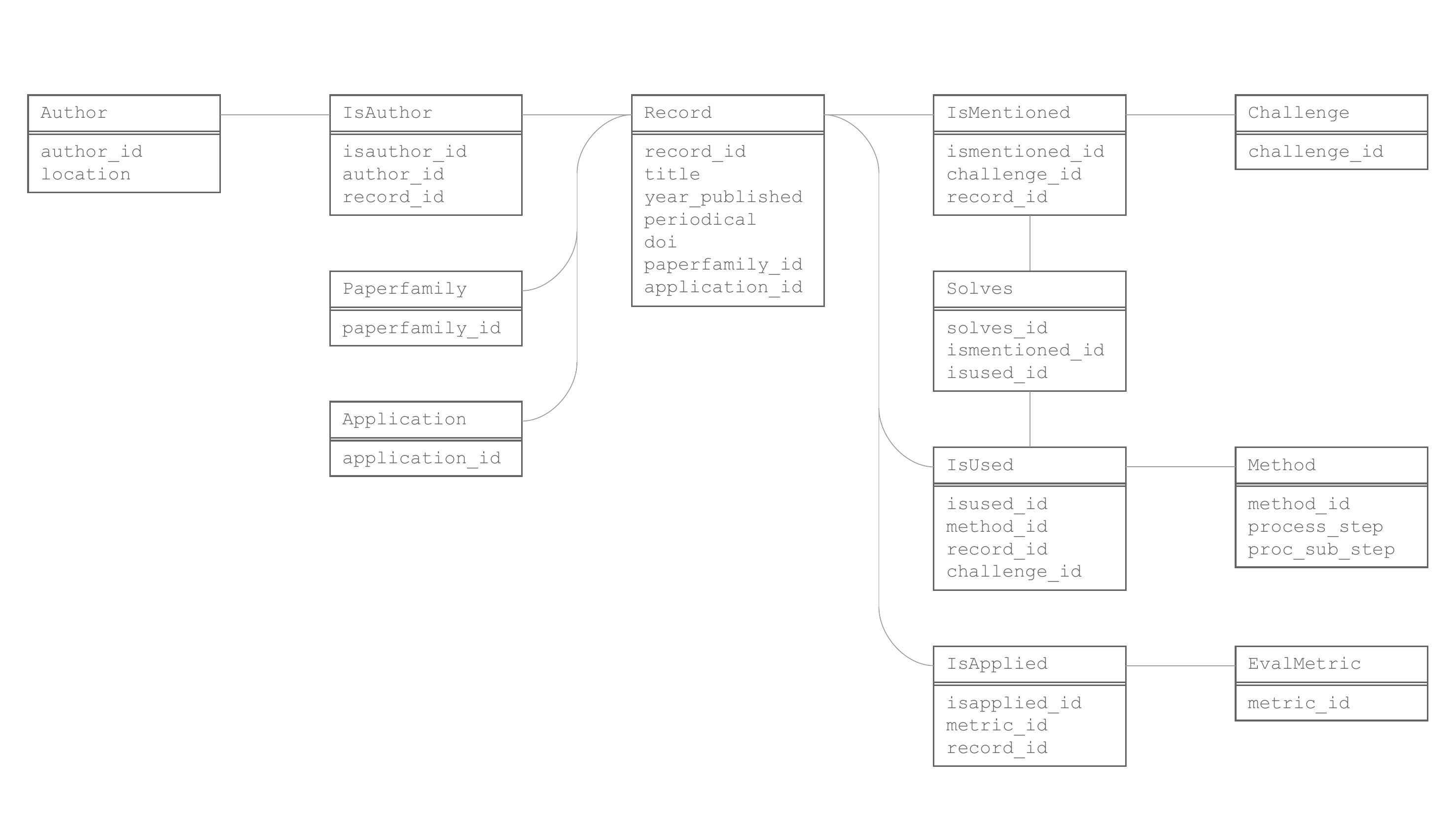}
    	\caption[Database design]{Database design}
    	\label{fig:Database}
    \end{figure}
    
    \begin{figure}[htbp]
    	\centering
    	\footnotesize
    	\includegraphics[width=0.75\textwidth]{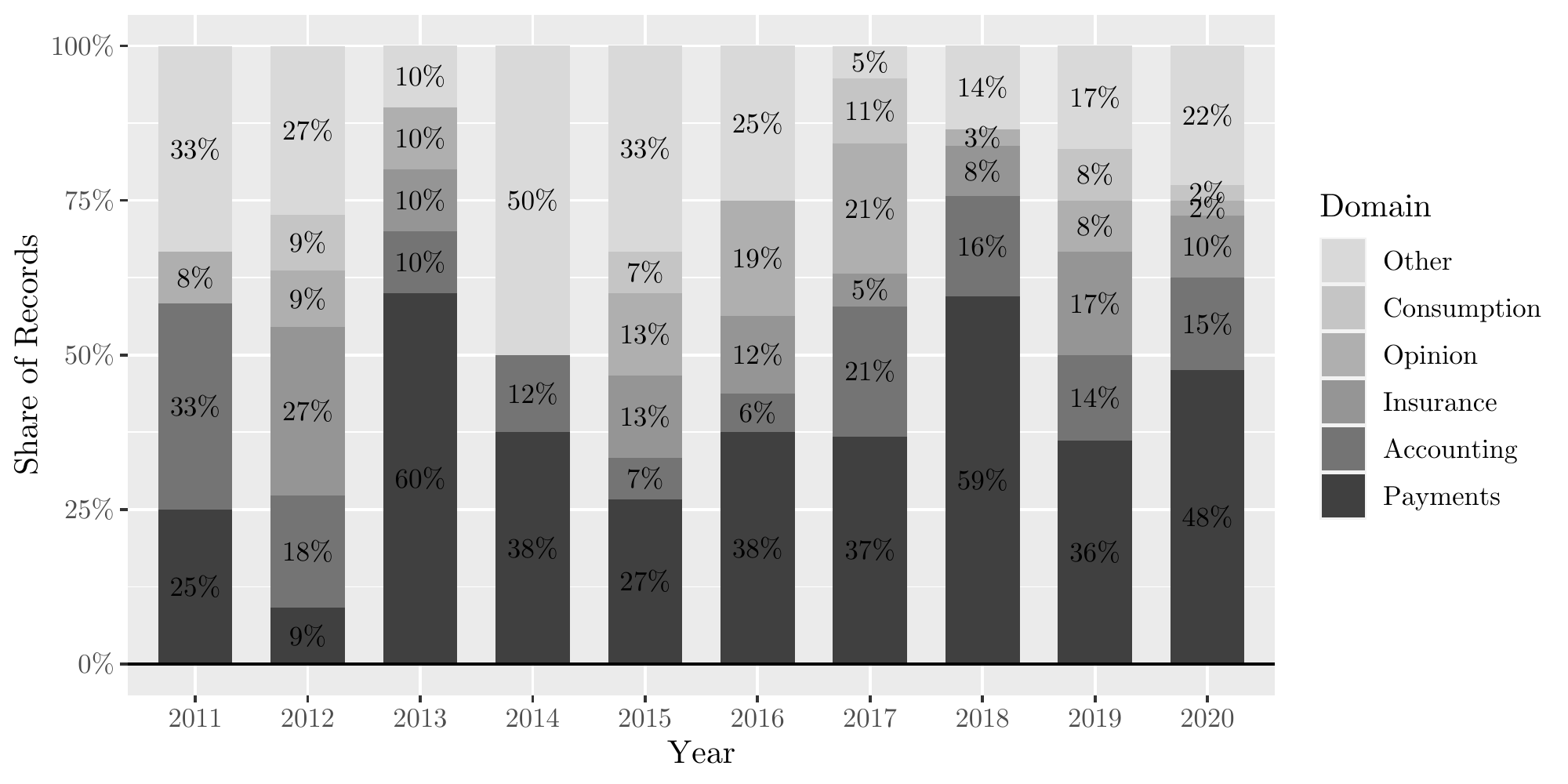}
    	\caption[Domains over time]{Domains over time}
    	\label{fig:Domains_over_time}
    \end{figure}
    
    \begin{figure}[htbp]
    	\centering
    	\footnotesize
    	\includegraphics[width=0.45\textwidth]{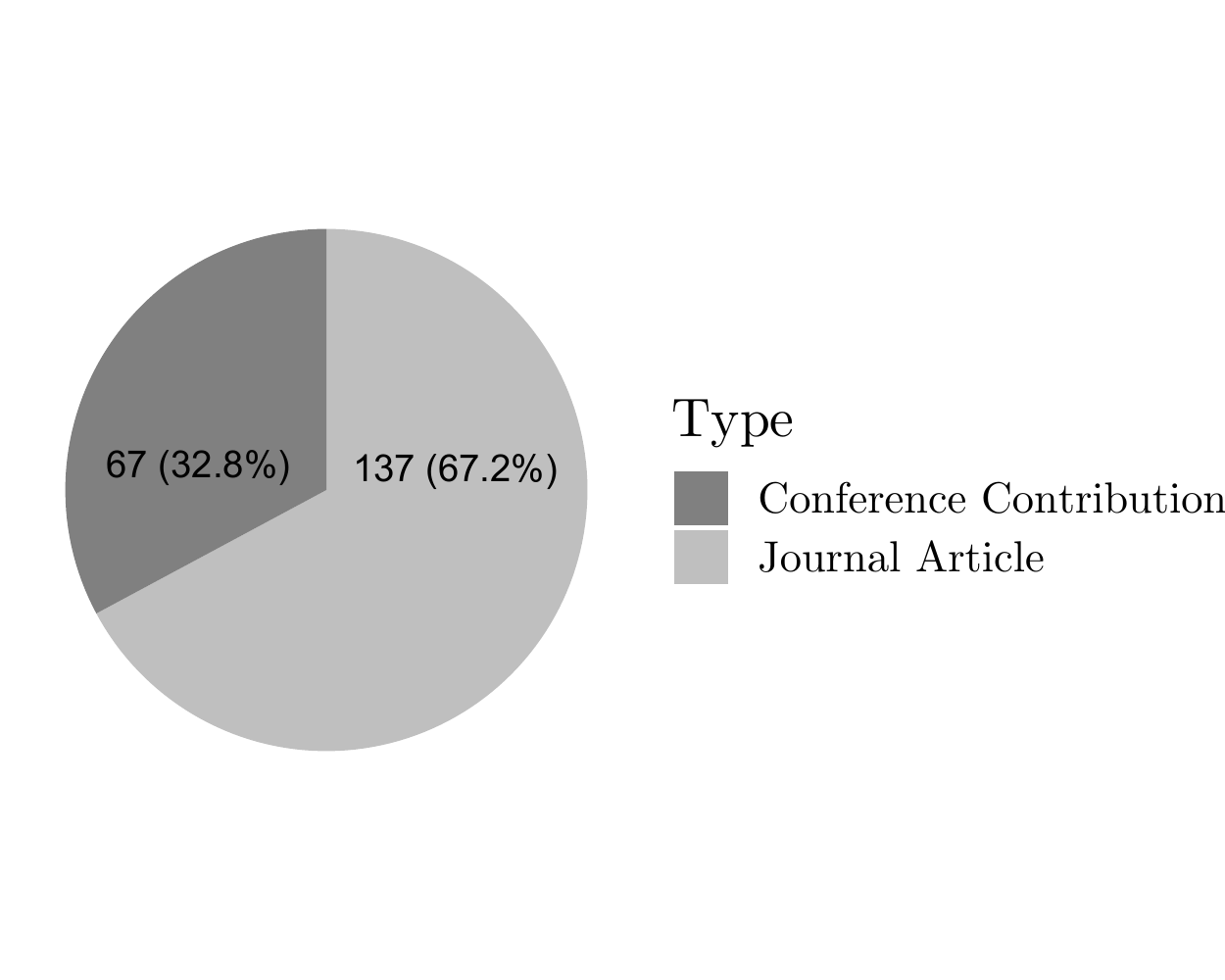}
    	\caption[Records per periodical]{Records per periodical}
    	\label{fig:Periodical}
    \end{figure}
    
    \begin{figure}[htbp]
    	\centering
    	\footnotesize
    	\includegraphics[width=0.45\textwidth]{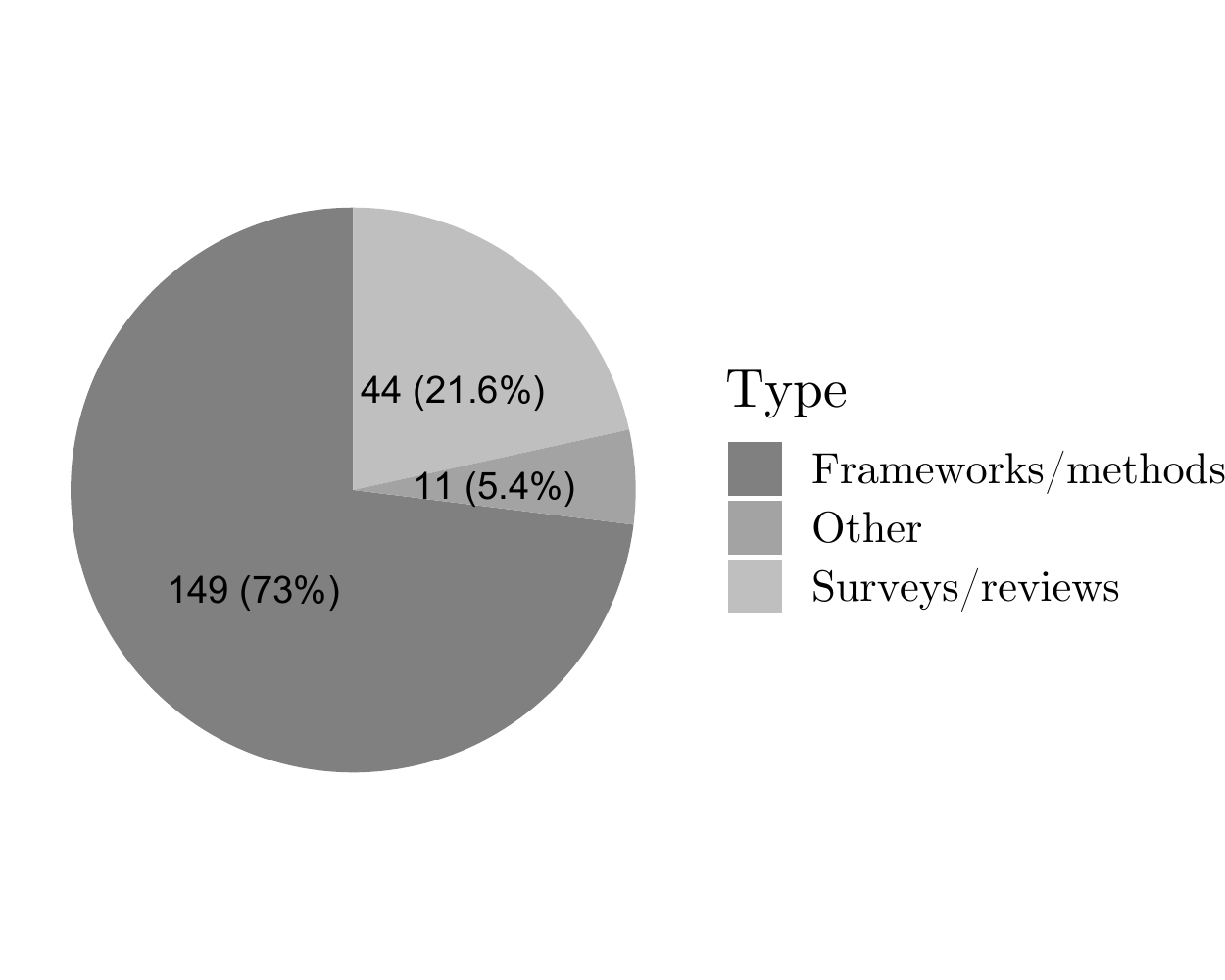}
    	\caption[Records per type]{Records per type}
    	\label{fig:Papertypes}
    \end{figure}
	\newpage

    \section{Record Selection}

    \printbibliography[keyword=record-selection, heading=none]
\end{appendix}
	
\end{document}